\documentclass[aps,twocolumn,letterpaper,amsmath,amssymb,floatfix,showpacs,superscriptaddress]{revtex4-1}

\usepackage{dcolumn}   
\usepackage{hyperref}
\usepackage{verbatim}
\usepackage{graphicx}
\usepackage{bm} 
\usepackage{dcolumn}  
\usepackage{epsfig}
\usepackage{color}

\begin{document}

\title{Global analysis of the ground-state wrapping conformation of a charged polymer on an oppositely charged nano-sphere}

\author{Hoda  Boroudjerdi}
\affiliation{Fachbereich Physik, Freie Universit\"at Berlin, Arnimalle 14, 14195 Berlin, Germany}

\author{Ali Naji}
\thanks{Corresponding author -- Electronic mail: \texttt{a.naji@ipm.ir}}
\affiliation{School of Physics, Institute for Research in Fundamental Sciences (IPM), Tehran 19395-5531, Iran }

\author{Roland R. Netz}
\affiliation{Fachbereich Physik, Freie Universit\"at Berlin, Arnimalle 14, 14195 Berlin, Germany}

\begin{abstract} 
 We investigate  the wrapping  conformations of a strongly adsorbed polymer chain on an oppositely charged nano-sphere by employing  a reduced (dimensionless) representation of a primitive chain-sphere model. This enables us to determine the global phase behavior of the chain conformation in a wide range of values for the system parameters including  the chain contour length, its linear charge density and persistence length as well as the nano-sphere charge and radius, and also the salt concentration in the bathing solution. The phase behavior of a charged chain-sphere complex can be described in terms of a few distinct conformational symmetry classes (phases) 
separated by continuous or discontinuous transition lines which are determined by means of appropriately defined (order) parameters. Our results can be applied to a wide class of strongly coupled polymer-sphere complexes including, for instances, complexes that comprise a mechanically flexible or semiflexible polymer chain or an extremely short or long chain and, as a special case, include the previously studied example of DNA-histone complexes. 
 \end{abstract}


\maketitle

\section{Introduction}

Charged polymers (or polyelectrolytes) can strongly bind to and wrap around oppositely charged spherical  
objects in aqueous  solutions. Such complexes are ubiquitous 
in soft matter physics, chemistry and nano-technology with many notable examples including complexes formed between synthetic polymers and  
globular proteins \cite{Strauss,Xia},  plastic beads \cite{Haronska}, latex \cite{Gana} and gold particles \cite{Gittins,Caruso}, charged micelles 
\cite{Dubin88,Li,Dubin90,Dubin1}, dendrimers \cite{Bielinska} and soft (polymer-grafted) nano-particles \cite{Cao2013,Cao2013b}. 
The size of these complexes can vary widely from just a few nanometers to several microns. 
Complexation of colloidal objects with charged polymers can strongly modify the effective interaction between these objects 
and thus further stabilize or destabilize colloidal suspensions \cite{Israelachvili,Polyelec}, leading to many important  
technological applications \cite{Mohwald,Rieger,Polyelec}. 

Charged polymer-sphere complexes are also very common in molecular biology and play an important role in intra-cellular processes \cite{The_cell}. The most remarkable
example is the packaging of DNA in the nucleus of eukaryotic cells, where a long (negatively charged) DNA chain is complexed with 
(positively charged) histone proteins forming the so-called {\em chromatin fiber} \cite{The_cell,Schiessel_chromatin,Schiessel03_rev,Kornberg,Luger1,Widom,Yao,Holde}.
The basic structural unit of chromatin is a highly charged polymer-protein complex known as the {\em nucleosome  core particle}, which is made of 
a DNA segment of 146 base pairs tightly wrapped (in nearly a 1-and-3/4  left-handed helical turn) around a cylindrical, wedge-shaped histone octamer. 
Both chromatin and nucleosome  core particles show striking conformational changes  with the concentration of additional salt in the solution, indicating  the
predominant  role played by the electrostatic interactions. 
(see, e.g., Refs. \cite{The_cell,Thoma,Allan,Gerchman,Woodcock,Bednar,Horowitz,hoda-chromatin}  and references therein for 
the salt-dependent behavior of chromatin which will not be discussed further in this paper). 
The importance of electrostatic effects in the case of nucleosome core particles is supported by a large body of {\em in vitro}
experimental observations that have been reviewed elsewhere (see, e.g., Refs. \cite{Kunze1,Kunze2,borujerdi,borujerdi2,borujerdi3,Messina04,Messina09,Schiessel03_rev,phys_rep,hoda-chromatin,thesis,Cherstvy11,Faraudo} and references therein). 
It has been found experimentally that upon decreasing the salt concentration 
the native {\em wrapped state} of DNA in nucleosome core particles (which is stable at intermediate salt concentrations about the physiological regime of 100~mM NaCl)
undergoes an unwrapping transition to an {\em expanded state}, where DNA is partially unwrapped from the histone core  \cite{Yager1,Yager2}. 

Recent analytical studies \cite{Muthu94,Sens99,Mateescu99,Park99,Netz99a,Kunze00,Nguyen01,Schiessel03,Vries06,Winkler05,Winkler06a,Winkler06b,Winkler07,Winkler11,Winkler12,Cherstvy11,Chervanyova,Wang11} and  numerical simulations \cite{Wallin96_I,Wallin96_II,Wallin97_III,Kong98,Jonsson01a,Jonsson01b,Stoll01a,Stoll01b,Stoll02,Stoll11,Akinchina02,Dzubiella03,Messina03,Maiti06,Carlsson01,Linse,Cao2013,Cao2013b,Faraudo} have provided very useful insight into the essential role that electrostatic interactions play in determining the structural properties of charged polymer-sphere complexes. It was shown that a minimal chain-sphere model for  nucleosome core particles \cite{Kunze1,Kunze2,borujerdi,borujerdi2,borujerdi3,Netz99a} can closely predict  the experimentally observed  
\cite{Dieterich, Ausio,Russev,Brown,Libertini1,Libertini2,Libertini3,Oohara,Weischet} salt-induced wrapping-unwrapping behavior of  these complexes based on an interplay between two competing mechanisms: 
the electrostatic attraction of DNA (modeled as a negatively charged worm-like chain) with the histone core (modeled as a positively charged nano-sphere),  
which favors DNA adsorption and wrapping around the sphere, and the mechanical bending stiffness of the DNA and its electrostatic self-repulsion, which favor
unwrapping and even dissociation of the DNA from the complex. 
The experimental stability diagrams \cite{Yager2} also exhibit an association (or complexation) equilibrium between  
nucleosome core  particles and the free DNA and free histones in an electrolyte solution in the regime of moderately large values of 
core particle and salt concentration. This behavior indicates that thermal chain fluctuations  \cite{Wallin96_I,Linse,Stoll02,Stoll11,Skepoe} become relevant 
at moderately large salt concentrations. These fluctuations can be accounted for  within the chain-sphere model as well  
and lead to a stability diagram that agrees qualitatively with the experimental one   \cite{borujerdi2}. 

From a theoretical point of view, charged polymer-sphere complexes present an interesting problem with a subtle interplay 
between elastic, electrostatic and possibly also entropic contributions from conformational changes of the polymer chain, giving rise to
to a diverse range of structural and thermodynamic properties. Rigorous analytical results, even within the simplest 
variations of the chain-sphere model, are still missing (except for the case of an uncharged polymer chain wrapping around cylindrical
inclusions where exact solutions have been found recently \cite{Hammant}) and most theoretical studies resort to approximate methods.
One can envisage two distinct limiting cases where well-defined approximations can be used. In the regime where 
thermal fluctuations are weak as compared with elastic or electrostatic contributions  (i.e., when the  energy scale of the chain adsorption on the sphere is large
and/or when the chain is highly charged or sufficiently stiff), one deals with {\em strongly coupled complexes}. These complexes can be 
studied by means of  ground-state-dominance approximations, strong electrostatic-correlations theories and other methods \cite{Netz99a,Kunze1,Kunze2,Nguyen1,Nguyen4,Nguyen21,Nguyen,Shklovs02,Marky2,Stoll01a,Stoll01b,Stoll02,Stoll11,borujerdi,borujerdi2,borujerdi3,Mateescu99,Jonsson01a,Jonsson01b,Schiessel,Schiessel1,Schiessel11,Sens,Skepoe,Wallin96_I,Linse}. They exhibit 
 striking properties such as a flat adsorbed polymer layer on the sphere with pronounced 
lateral order due to mutual electrostatic repulsions of the chain segments and an effective charge inversion of the sphere  
by an excess adsorption of polymer charge larger than necessary to neutralize its bare charge \cite{Mateescu99,Schiessel,Kunze1,Kunze2,Nguyen1,Park99,Nguyen4,Nguyen21,Nguyen,Shklovs02,Skepoe}. 
The {\em weakly coupled complexes}, on the other hand, exhibit weak polymer 
adsorption and thus strong chain fluctuations which can be treated, for instance, 
using field-theoretic methods  and give rise to a diffuse polymer layer bound around the sphere (see, e.g.,   \cite{Podgornik3,Sens99,Golestanian,Sens,Stoll01b,Stoll02,Stoll11,Muthu94} and references therein).   
There is another distinct regime of parameters where the charged polymer takes a rosette-shaped structure  with multiple-point
contacts with the sphere and large low-curvature loops connecting them  \cite{Schiessel03_rev,Schiessel1,Schiessel11,Akinchina02}. This regime is obtained
  for large polymer persistence length as compared with the sphere radius, large polymer  length and relatively small chain-sphere 
adsorption energy (not large enough to bring the whole chain into contact with the sphere). 
The analytical methods used to study these different regimes have been complemented by numerical simulations that  provide
a more comprehensive analysis of the properties of
charged polymer-sphere complexes within these regimes and beyond them \cite{Wallin96_I,Wallin96_II,Wallin97_III,Kong98,Jonsson01a,Jonsson01b,Stoll01a,Stoll01b,Stoll02,Stoll11,Akinchina02,Dzubiella03,Messina03,Maiti06,Carlsson01,Linse}.

In this paper, we shall primarily focus on strongly coupled chain-sphere complexes formed by a charged chain wrapped 
around an oppositely charged nano-sphere and explore the wrapping phase behavior of the
chain in a wide range of system parameters including the contour length, linear charge density, and
the persistence length of the chain as well as the sphere charge and radius, and also the salt concentration in the bathing solution.
Our main objective is to establish structural phase diagrams using a reduced (or dimensionless) description that can be representative of 
the global (or universal)  behavior of the system based on only a few {\em dimensionless} parameters (for example, the ratio
of the chain length to the sphere radius). This is achieved by  employing a primitive chain-sphere model that, as noted above, has proved  useful in the context of 
nucleosome core particles  \cite{Kunze1,Kunze2,borujerdi,borujerdi2,borujerdi3,Netz99a}; its main advantage being its simplicity that 
enables a systematic investigation of various aspects of charged polymer-sphere complexes especially when specific effects (such as histone tails \cite{Schiessel03_rev}) are not dominant and the generic properties of these complexes are of interest. 
We shall make use of a previously established numerical optimization
method for strongly coupled complexes \cite{borujerdi,borujerdi2,borujerdi3} in order to obtain the wrapping conformation of the chain and its symmetry classes (phases)
in various parts of the parameter space. We shall in particular study the case of  a mechanically flexible chain with no mechanical bending rigidity;
this latter quantity stiffens the chain and favors unwrapping of the chain from the sphere, and thus competes with 
the electrostatic self-repulsion of the chain segments, which does the same. Our analysis for a mechanically flexible chain thus brings out the role electrostatic 
mechanisms at work in the wrapping-unwrapping behavior of the chain more clearly and may be applicable to highly charged chains that have a small
mechanical (or bare) persistence length (such as, e.g., Polystyrene Sulfonate, PSS, which is an essentially flexible polymer with a linear charge density of up to 4 elementary charges per nanometer). The role of salt screening effects and the nano-sphere charge as well as the chain contour length and its persistence
length is discussed in detail as well. 

Our study therefore extends the previous works \cite{Kunze1,Kunze2,borujerdi3} that are specifically focused on the case of DNA-histone complexes   
to a wider class of polyelectrolyte-macroion complexes such as complexes involving a  mechanically flexible   chain  or an arbitrary chain  length 
relative to the sphere radius. These cases can be realized  by synthetic polyelectrolytes complexed with charged globular objects of different 
sizes such as plastic or latex beads and colloids \cite{Haronska,Gana,Mohwald,Xia}, micelles \cite{Dubin88,Li,Dubin90,Dubin1} as well as  nanoscopic 
gold particles \cite{Gittins,Caruso}. 

The organization of the paper is as follows: In Section \ref{sec:model}, we shall introduce our model and the numerical methods which we shall employ in 
our study, and in Section \ref{sec:upd_phase_behavior_nosalt}, we present an extensive analysis of our results for the case of charged chain-sphere complexes
in the absence of salt screening effects, which will be summarized in a universal phase diagram in terms of the rescaled chain length and
the rescaled sphere charge. In Section \ref{sec:phase_behavior_salt}, we investigate the effects of salt screening  on the wrapping state 
and conformational changes of the polymer  around the oppositely charged sphere and finally in Section \ref{sec:phase_behavior_ellp}, we turn
to the effects the of the mechanical persistence length of the chain. We shall conclude our discussion in Section \ref{sec:discussion}. 

\section{Model and Methods}
\label{sec:model}

We model the nano-sphere as a uniformly charged sphere of radius $R_{\mathrm{s}}$ and total charge  $Ze_0$, where $e_0$ is the elementary charge 
and, with no loss of generality, we assume  $Z>0$. The oppositely charged polymer is modeled as an inextensible 
(worm-like) chain of contour length $L$, bare mechanical persistence length $l_{\mathrm{p}}$,
and uniform linear charge density $-\tau e_0$, where we take $\tau>0$. 
We shall later consider both mechanically flexible ($l_{\mathrm{p}}=0$) and semi-flexible chains ($l_{\mathrm{p}}>0$). Note that the radius of the sphere is taken 
such that it accounts for the finite radius of the polymer as well, therefore, $R_{\mathrm{s}}$ represents the minimal distance between the sphere center and the centers of the chain's monomers. 
We shall furthermore assume that the system is immersed 
in a background monovalent electrolyte of bulk concentration $n_s$ that screens the bare Coulomb interactions of the charges  on the nano-sphere and the polymer by introducing
an inverse Debye screening length (or salt screening parameter)  of $\kappa = \sqrt{8\pi \ell_{\mathrm{B}} n_s}$, where $\ell_{\mathrm{B}} = e_0^2/(4\pi \varepsilon \varepsilon_0 k_{\mathrm{B}}T)$ is the Bjerrum length. In water (with the
dielectric constant $\varepsilon\simeq 80$) and at room temperature ($T=300$K), we have $\ell_{\mathrm{B}} \simeq 0.7$~nm and thus under   physiological conditions (with $n_s\simeq 100$~mM), we obtain $\kappa^{-1}\simeq 1$~nm. The electrostatic interactions between polymer/sphere charges is described by means of a Debye-H\"uckel (DH) pair potential which is expected to be valid for monovalent electrolytes of sufficiently large concentration (see the Discussion and Refs. \cite{Kunze1,Kunze2,borujerdi3,thesis,phys_rep} for further details relating to the validity regime of the DH approximation).  

The Hamiltonian of the above chain-sphere model for a given chain configuration parameterized by ${\mathbf r}(s)$ (with $0\leq s\leq L$) can be written as 
\begin{eqnarray}\label{eq:upd_H_full}
{\mathcal H}[  {\mathbf r}(s)] &=& 
             \frac{l_{\mathrm{p}}}{2  }\int_0^{  L}{\mathrm{d}}  s\,\,\big(\ddot {  {\mathbf r}}(  s)\big)^2 + \\            
		&&\hspace{-1.2cm}+\,  \tau^2\ell_{\mathrm{B}}\int_0^{  L}{\mathrm{d}}  s\int_{  s}^{  L}{\mathrm{d}}  s'\,\,
       			\frac{e^{-  \kappa |  {\mathbf r}(  s)-  {\mathbf r}(  s')|}}
					{|  {\mathbf r}(  s)-  {\mathbf r}(  s')|}\nonumber \\
	        &&\hspace{-1.2cm}-\,\frac{ Z \tau \ell_{\mathrm{B}} }{1+  \kappa   R_{\mathrm{s}}} \int_0^{  L}{\mathrm{d}}  s \bigg[
	            \frac{e^{-  \kappa  (|  {\mathbf r}(  s)|-  R_{\mathrm{s}})}}{|  {\mathbf r}(  s)|}
	              -   A\, e^{-(|  {\mathbf r}(  s)|-  R_{\mathrm{s}})/  \alpha}\bigg]. \nonumber
\end{eqnarray}
The first term in this equation represents the contribution from  the mechanical bending rigidity of the chain, the second term gives the electrostatic self-energy of the chain due to 
the inter-repulsion between chain segments, and the  third term gives the chain-sphere interactions, including the electrostatic attraction energy 
of the chain with the nano-sphere (first term in the brackets) and a semi-rigid excluded-volume repulsion (second term in the brackets). 
The latter is characterized by two parameters $A$ and $\alpha$ for the strength and the range of the excluded-volume repulsion, respectively. In order to guarantee the
impenetrability of the sphere, we choose  $A\ll 1/R_{\mathrm{s}}$ and $\alpha<\kappa^{-1}$ (in most cases, we take  $A= 0.014$~nm$^{-1}$ and $\alpha= 0.02$~nm 
in order to obtain an equilibrium sphere-monomer separation that is equal to the sphere radius $R_{\mathrm{s}}$ within 2\%).

\subsection{Dimensionless representation}
\label{sec:upd_d_less}

The above expression for the Hamiltonian introduces several different parameters which all may affect the 
structure of a complex. The space of parameters can be spanned  by a few dimensionless parameters,
which can be determined by, e.g., rescaling the coordinates with a given length scale such as the sphere radius $R_{\mathrm{s}}$ as $\tilde s  = {s}/{R_{\mathrm{s}}}$
and $\tilde {\mathbf r}(\tilde s)  = {\mathbf r}(s)/{R_{\mathrm{s}}}$. The dimensionless representation for the  Hamiltonian thus follows as
\begin{eqnarray}\label{eq:H_full_rescale_R}
\tilde {\mathcal H}[\tilde {\mathbf r}(\tilde s)] &=& 
             \frac{\tilde l_{\mathrm{p}}}{2 }\int_0^{\tilde L}{\mathrm{d}}\tilde s\,\,\big(\ddot {\tilde {\mathbf r}}(\tilde s)\big)^2 + \\    
		&&\hspace{-1.2cm}+\, \int_0^{\tilde L}{\mathrm{d}}\tilde s\int_{\tilde s}^{\tilde L}{\mathrm{d}}\tilde s'\,\,
       			\frac{e^{-\tilde \kappa |\tilde {\mathbf r}(\tilde s)-\tilde {\mathbf r}(\tilde s')|}}
					{|\tilde {\mathbf r}(\tilde s)-\tilde {\mathbf r}(\tilde s')|}\nonumber \\
	        &&\hspace{-1.2cm}-\, \frac{ \tilde Z}{1+\tilde \kappa } \int_0^{\tilde L}{\mathrm{d}}\tilde s \bigg[
	            \frac{e^{-\tilde \kappa  (|\tilde {\mathbf r}(\tilde s)|-1)}}{|\tilde {\mathbf r}(\tilde s)|}
	              - \tilde A\, e^{-(|\tilde {\mathbf r}(\tilde s)|-1)/\tilde \alpha}\bigg],\nonumber
\end{eqnarray}
where $\tilde {\mathcal H} =  {\mathcal H}/ (\tau^2 R_{\mathrm{s}}  \ell_{\mathrm{B}}  k_{\mathrm{B}} T)$, and
the key dimensionless parameters are then obtained as 
\begin{equation}
   \tilde l_{\mathrm{p}} = \frac{l_{\mathrm{p}}}{\tau^2 R_{\mathrm{s}}^2  \ell_{\mathrm{B}} }, \,\,\,\,\,
   \tilde Z = \frac{Z}{\tau R_{\mathrm{s}} },\,\,\,\,\,
   \tilde \kappa = \kappa R_{\mathrm{s}},\,\,\,\,\,
   \tilde L  = \frac{L}{R_{\mathrm{s}}}.
\end{equation}
These parameters correspond to the rescaled bare persistence length, the rescaled  sphere charge, the rescaled inverse Debye screening
length and the rescaled chain contour length, respectively. (The parameters associated with the excluded-volume repulsion are rescaled
as $\tilde \alpha =  \alpha/R_{\mathrm{s}}$ and $\tilde A =  A R_{\mathrm{s}}$, which are fixed to some appropriate
values as noted before and will not be considered as control parameters). 
  
\begin{figure*}[t]
\begin{center}
\includegraphics[angle=0,width=14.cm]{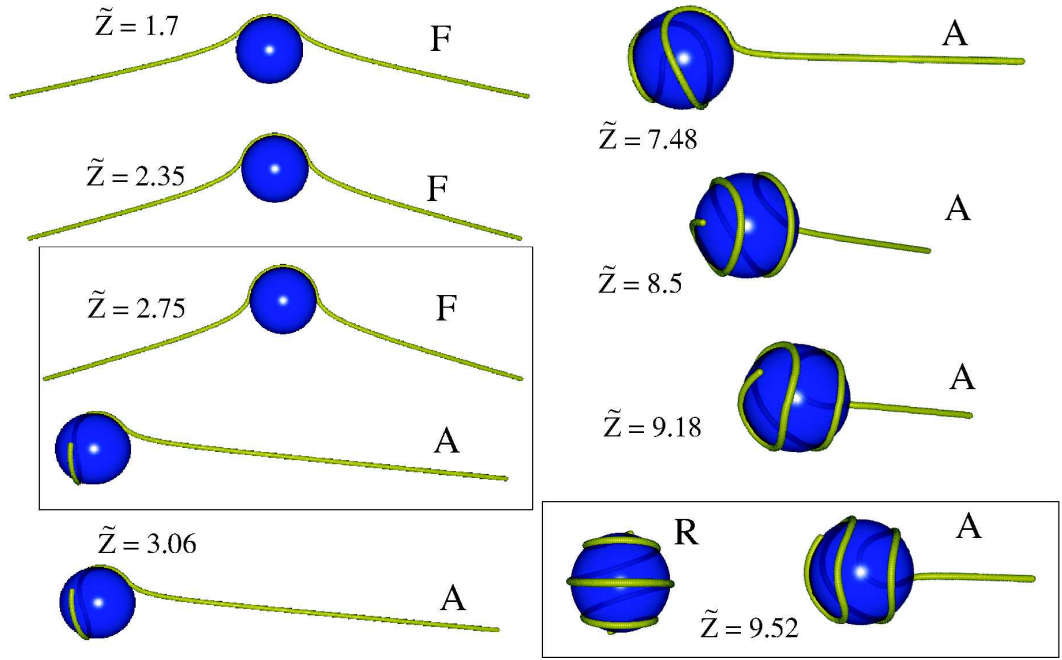}
\caption{Optimal conformations of a charged chain around an oppositely charged  nano-sphere obtained by numerical minimization in the case of a mechanically flexible chain ($\tilde l_{\mathrm{p}} = 0$) 
with rescaled contour length 
$\tilde L = L/R_{\mathrm{s}} = 16.3$ in the absence of salt screening ($\tilde \kappa = 0$) for various rescaled 
sphere charge values $\tilde Z$  as shown on the graph. The symmetry class of the chain conformation 
(F, A or R) is also indicated for each configuration. The coexisting states at the transition point between phases F and A, $\tilde Z = 2.75$,  
and also at the transition point between phases A and R, $\tilde Z =9.54$, are shown together in a closed box. For the sake of presentation, the configurations on the right are  enlarged to show the details more clearly.}
\label{fig:upd_zlgal}
\end{center}
\end{figure*}
  
\subsection{Numerical optimization}
 
We shall make use of a numerical minimization scheme that has been employed extensively in previous studies of DNA-histone complexes 
\cite{Kunze1,Kunze2,borujerdi,borujerdi2,borujerdi3,hoda-chromatin,thesis}. The full details of this scheme can be found in these references and 
will not be reiterated here and thus we shall delimit our discussion to only a few of the key conceptual elements of our approach. 
 
It should be noted first that in the case of  strongly coupled chain-sphere complexes, that are of concern in our study, the strength of electrostatic interactions 
is typically large as compared with the  thermal energy and, hence, the electrostatic self-repulsion of the chain segments is large 
and/or the chain is strongly adsorbed onto the sphere due to the strong chain-sphere attraction. The electrostatic 
adsorption energy of the chain can indeed be very large even for modestly large values of chain/sphere charges across a whole range of salt concentration
including the physiologically relevant regime as is the case, for instance,  for the DNA-histone system \cite{Kunze1,Kunze2,Schiessel03_rev}. 
The thermal fluctuations can be negligible also when the chain has a large mechanical persistence length; this is important  especially when electrostatic 
stiffening of the chain due its self-repulsion is screened out by the additional salt. In these cases, the spatial conformation of the chain wrapped 
around the sphere can be obtained by minimization  the Hamiltonian of the system
with respect to the chain degrees of freedom. This corresponds to the so-called {\em ground-state-dominance} approximation and yields energetically
optimal (ground-state) configurations of the chain-sphere complex. 

The aforementioned optimization procedure can be done most efficiently by  
numerical methods such as the quasi-Newton numerical minimization algorithm on a properly discretized model of the chain \cite{Kunze1,Kunze2,borujerdi,borujerdi2,borujerdi3,hoda-chromatin,thesis}. In most cases, the symmetries of the numerically
obtained optimal configurations already indicate whether the result corresponds to the global minimum of the Hamiltonian but further 
checks (including a combination of parameter quenching and stochastic perturbation methods) 
must be done to ensure the global stability of the optimal solutions. This becomes important especially near   discontinuous transition points, 
where several optimal states of the same energy coexist and thus, as will be shown in a few examples in the following Section, the system may exhibit 
meta-stable states with energies close to the ground-state configuration.

\section{Conformational transitions in the no-salt limit}
\label{sec:upd_phase_behavior_nosalt}

Let us first consider the case of a charged complex involving a {\em mechanically} flexible chain with  no mechanical persistence length ($l_{\mathrm{p}}=0$) and in the absence of salt screening ($\kappa=0$). In this case, the long-ranged (unscreened) electrostatic self-repulsion 
of the chain segments still generate an effective electrostatic stiffness  \cite{netz-orland,Skolnick77,Odijk,Fixman,Odijk95,Khokhlov}) but the problem reduces to 
a slightly simpler version as,  in the absence of a mechanical bending rigidity, the electrostatic self-repulsion of the polymer  is the only mechanism that opposes its wrapping around the nano-sphere. The effects due to of salt screening and mechanical persistence length will be investigated later. 

\subsection{Role of the nano-sphere charge}
\label{sec:upd_effZ}

In Fig. \ref{fig:upd_zlgal}, we show the polymer conformations as obtained by minimizing the Hamiltonian, Eq. 
(\ref{eq:H_full_rescale_R}), for a number of different values for the sphere charge and for fixed  rescaled chain length of $\tilde L =16.3$ (for a sphere diameter of $R_{\mathrm{s}}=5$~nm,  we have an actual chain contour length of $L=81.6$~nm comparable to 240 base pairs of DNA). 
If the sphere charge is set to zero, the chain takes the  conformation of a straight line in its ground state (not shown). For non-zero but sufficiently small sphere charge (e.g.,  $\tilde Z=1.7$ and $\tilde Z=2.35$ in Fig. \ref{fig:upd_zlgal}), 
a finite length of the chain is partially wrapped around the sphere and thus the continuous rotational symmetry of the free (straight-line) conformation  is broken and reduces to a {\em two-fold rotational symmetry} with respect to the axis connecting the chain mid-point to the sphere center. 
In this case, the chain conformation is also planar and  {\em mirror symmetric}. These symmetries can be determined
by means of appropriately defined orders parameters \cite{Kunze1,Kunze2,borujerdi3,thesis} which we shall discuss later in Section \ref{sec:upd_ZLphase}. 
The regime of parameters where the ground-state  conformation of the chain has  both the two-fold rotational symmetry and the mirror symmetry is 
referred to as the {\em expanded} or {\em fully symmetric phase} F. Clearly, in this phase the long-range self-repulsion of chain segments is dominant, but 
as the sphere charge is increased, the chain is wrapped more strongly around the nano-sphere and its conformational shape 
undergoes a number of structural phase transitions.

For the rescaled sphere charge $\tilde Z = 2.75$, 
the sphere-chain attraction becomes strong enough to break simultaneously both the rotational and mirror 
symmetries. At this point, the expanded configuration (phase F) {\em coexists} 
with an {\em asymmetric} three-dimensional configuration, referred to as phase A, that are obtained as  
optimal configurations with equal energies (shown together in a closed box). This value of the sphere charge therefore represents the locus of a discontinuous
transition that can be determined from the changes in the  energy of the chain-polymer complex (see below) or the corresponding order parameters (Section \ref{sec:upd_ZLphase}). Note that in the asymmetric phase A, the chain wraps around the sphere from one of 
its end points, and that for a nano-sphere of  radius $R_{\mathrm{s}}=5$~nm, the threshold charge $\tilde Z = 2.75$ is obtained by 
taking a polymer
linear charge density of $\tau = 5.88$~nm$^{-1}$ (which is appropriate for DNA) and  the actual sphere charge $Z= 81$.

By increasing the sphere charge further, the degree of chain wrapping  is continuously  increased, while the complex remains
in the asymmetric phase A with an extended arm stretching out from one of the polymer ends.
A second transition occurs at $\tilde Z = 9.52$, where the chain-sphere attraction becomes
large enough to completely wrap the chain around the sphere, which, for the parameters in the figure, 
exhibits nearly three complete turns. This transition thus restores the  two-fold rotational symmetry and 
gives rise to a {\em completely wrapped} state R. It is again a discontinuous transition where the two states A and R coexist as shown in a closed box in the figure.

\begin{figure}[t]
\begin{center}
\includegraphics[angle=0,width=8.5cm]{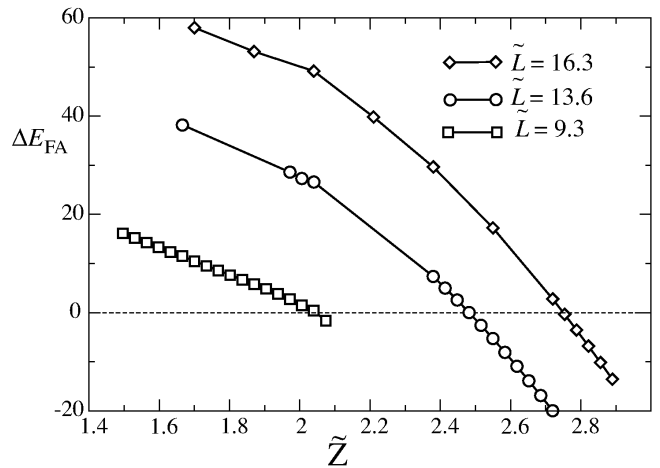}
\caption{Energy difference between the optimal configuration in  phases F and A,
$\Delta E_{\mathrm{FA}} = E_{\mathrm{A}} - E_{\mathrm{F}}$, is shown 
as a function of $\tilde Z$ for the case with $\tilde l_{\mathrm{p}} = 0$ and $\tilde \kappa = 0$ 
and different rescaled chain contour lengths $\tilde L = 16.3$ (corresponding to the configurations shown
in Figure \ref{fig:upd_zlgal}), $\tilde L = 13.6$  and $\tilde L = 9.3$.}
\label{fig:upd_tran}
\end{center}
\end{figure}

 \begin{figure*}[tbh]
\begin{center}
\includegraphics[angle=0,width=14cm]{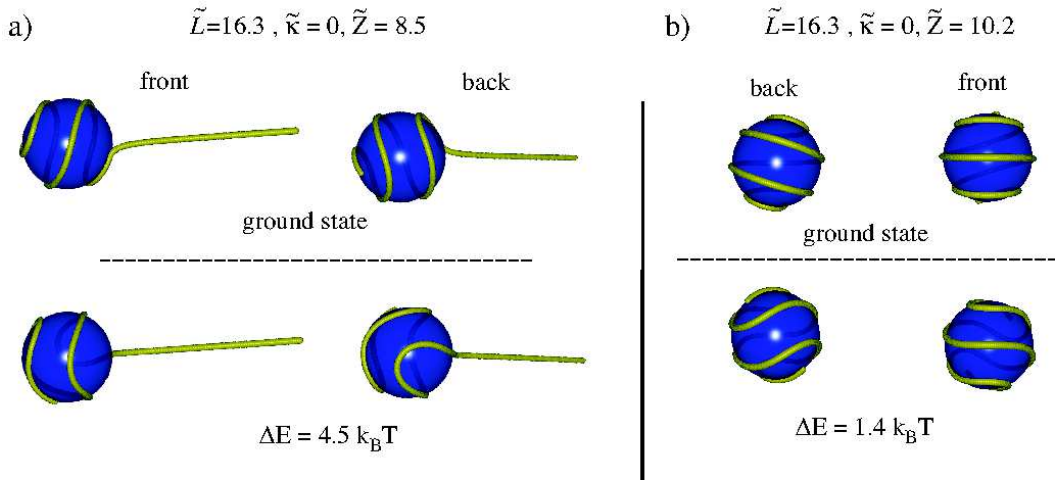}
\caption{Typical meta-stable configurations (bottom) in  the symmetry classes A (panel a) and R (panel b) shown from both back and front views and along with 
their corresponding ground-state configurations (top). Here we have $\tilde l_{\mathrm{p}} = 0$ and $\tilde \kappa =0$. The values of  other parameter  as well as
the energy difference between each of these meta-stable states and the ground state, $\Delta E$,  are shown on the graph.}
\label{fig:upd_meta}
\end{center}
\end{figure*}

\begin{figure*}[t]
\begin{center}
\includegraphics[angle=0,width=16.cm]{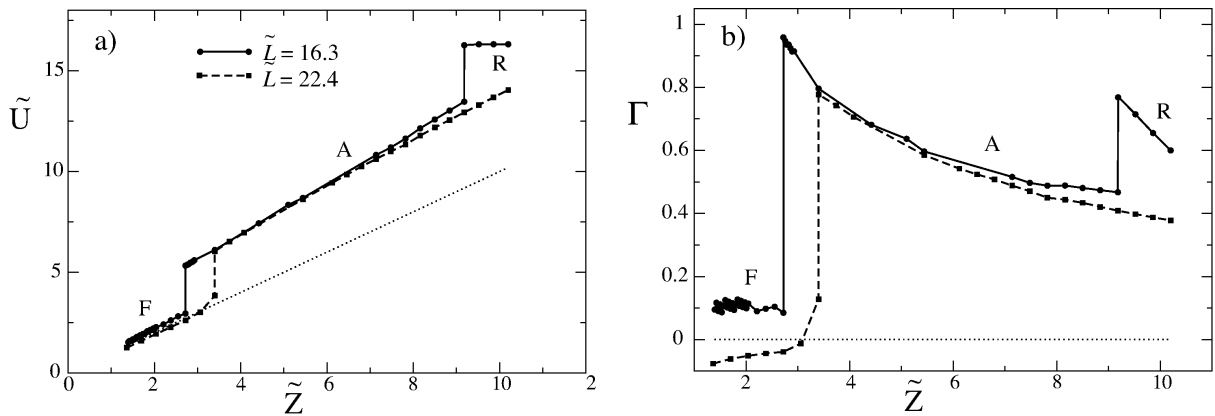}
\caption{a) Total polymer charge adsorbed on the sphere in rescaled units,  $\tilde U$ (see the text) and b) the  degree of charge inversion, $\Gamma$, Eq. (\ref{eq:upd_gamma}), are plotted
as a function of the rescaled sphere charge $\tilde Z$ for $\tilde l_{\mathrm{p}} = 0$ and $\tilde \kappa = 0$ and two different rescaled chain lengths 
$\tilde L = 16.32$ and 22.4. The thin dotted line shows the isoelectric line with  $\tilde U=\tilde Z$.}
\label{fig:upd_overZ}
\end{center}
\end{figure*}

As noted above, the coexisting states correspond to different minima of the Hamiltonian with equal energies at the transition
point. As the control parameter $\tilde Z$ is varied across the transition point, only one of these minima remains stable 
and the other one assumes a higher energy and can thus persist only as a  {\em meta-stable} state; it may eventually disappear as a local 
minimum of the Hamiltonian. For instance, in Fig. \ref{fig:upd_zlgal}, phase F persists as a meta-stable state when
the sphere charge increases beyond the discontinuous transition point to phase A at $\tilde Z = 2.75$. One can therefore 
determine the locus of the discontinuous
transition using the energy difference $\Delta E_{\mathrm{FA}} \equiv E_{\mathrm{A}} - E_{\mathrm{F}}$, where
$E_{\mathrm{F}}$  and  $E_{\mathrm{A}}$ are the energies of the numerically determined configurations A and F. 
This is shown in Fig. \ref{fig:upd_tran} for three different chain contour lengths, which clearly show that phase A takes over and
becomes  stable  beyond the transition point  ($\Delta E_{\mathrm{FA}} =0$) and that both states can be actually present as  meta-stable states
for a wide range of values for $\tilde Z$. Also, the transition point appears to show a strong dependence on the chain length and shifts to smaller 
$\tilde Z$ values as $\tilde L$ is decreased, which means that  for shorter chains, phase A becomes stable in a wider range 
of rescaled sphere charges. For a sufficiently small contour length comparable to  
 $L\simeq 2\pi R_{\mathrm{s}}$, the asymmetric phase  A completely disappears and
the fully symmetric phase, F,  turns out to be the only solution for the ground-state configuration.
The influence of the chain length on its wrapping behavior  will be explored further in the following Sections.

At this point, it is useful to note that in addition to the meta-stable states encountered at 
the discontinuous transition points, there may be  additional meta-stable states with energies very close to
the ground-state configurations that may or may not belong to any of the ground-state symmetry classes for any given set 
of parameters. Such meta-stable states  typically show more enhanced
chain undulations on the nano-sphere as compared with the ground-state conformation. 
This indicates that thermal fluctuations may become important where such excited states are relevant and 
may induce transitions between these locally stable states. In Fig. \ref{fig:upd_meta}, we show
two such meta-stable states (bottom) for  symmetry classes A and R and compare them with 
their corresponding ground-state configurations (top).  As indicated on the graph, the energy difference from the ground state energy
can be as low as a few $k_{\mathrm{B}}T$, e.g., for the example shown for phase R in Fig. \ref{fig:upd_meta}b, we have $\Delta E = 1.4 k_{\mathrm{B}}T$. 

 \begin{figure*}[t]
\begin{center}
\includegraphics[angle=0,width=15.cm]{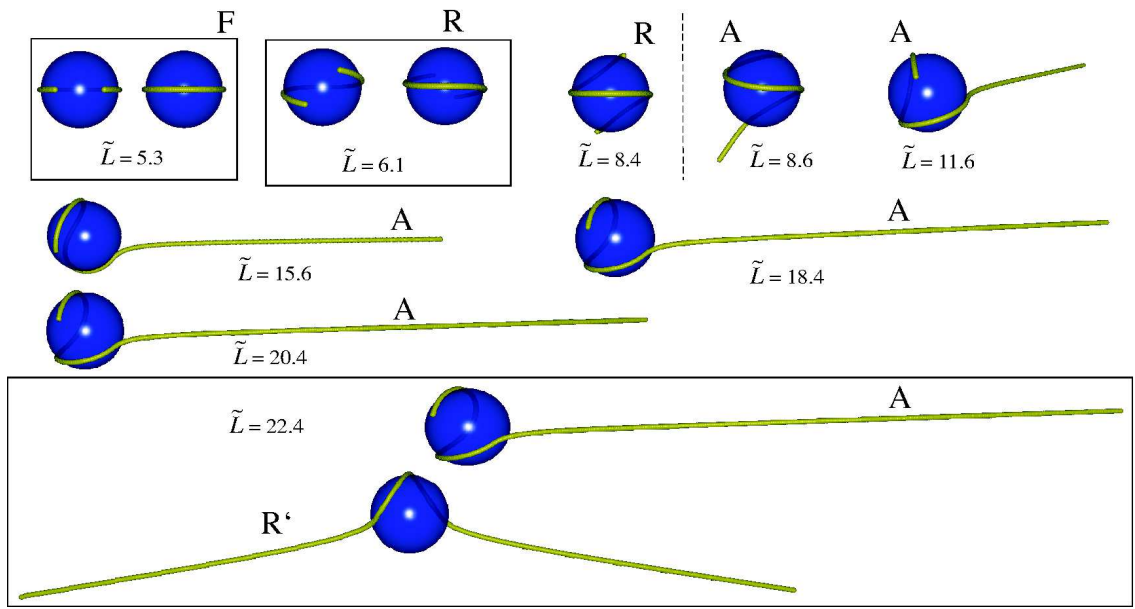}
\caption{Chain wrapping conformations  for fixed $\tilde Z = 3.4$, $\tilde l_{\mathrm{p}} = 0$ and $\tilde \kappa = 0$ and different
rescaled chain length $\tilde L$ increasing from $\tilde L=5.3$ up to $\tilde L=22.4$. The symmetry class of the chain conformation 
(F, A, R, or R$'$) is also indicated for each configuration. 
For $\tilde L=5.3$ and $\tilde L=6.1$ (top left), we show both the front and back views of the same configuration together 
in a closed box. The  configurations for $\tilde L=22.4$ (bottom) show the two coexisting states at the transition point from phase F to R$'$.}
\label{fig:upd_Lgal}
\end{center}
\end{figure*}

\subsection{Overcharging of the nano-sphere}

The  wrapping transitions discussed in the preceding Section give rise to highly charge-inverted chain-sphere complexes where the sphere charge
is overcompensated by the amount of polymer charge adsorbed on the sphere.  This latter quantity can be defined
in terms of the length of polymer which is wrapped around the sphere,  $L_{\mathrm {w}}$ (conventionally, a chain bead is considered 
wrapped or adsorbed on the sphere if its relative distance from the sphere center is within 2\% of the sphere radius). Hence, the total
polymer charge adsorbed on the sphere is $U = \tau L_{\mathrm {w}}$ (in units of the elementary charge $e_0$), or in rescaled 
representation $\tilde U \equiv \tau L_{\mathrm {w}}/ \tau R_{\mathrm{s}} = \tilde L_{\mathrm {w}}$. The effective (net)
charge of the chain-sphere complex is then equal to $U-Z$, and hence, the degree of charge inversion can be defined as the ratio between
the net charge of the sphere and its bare charge as
\begin{equation}
\label{eq:upd_gamma}
  \Gamma = \frac{\tilde U}{\tilde Z} -1.
\end{equation}

In Fig. \ref{fig:upd_overZ},  we show our results for $\tilde U$ (panel a) and $\Gamma$ (panel b) as 
a function of the rescaled sphere charge valency $\tilde Z$. We have plotted
the results for two different cases with $\tilde L =16.3$ and $\tilde L = 22.4$ (corresponding to the length of about 240 and 330  base pairs of DNA
in actual units, respectively, provided that $R_{\mathrm{s}}=5$~nm). The thin dotted line in the figure 
shows the isoelectric line  $\tilde U=\tilde Z$, where the net charge of the complex is zero. 
It is clear that, for most of the range plotted
in the figure, the sphere charge is overcompensated by the polymer $\Gamma>0$ and thus, the net charge of the sphere changes sign relative
to its bare value. Also the transitions from phase F to A and from phase  A to R lead to discontinuous jumps in 
the magnitude of the net charge and consequently also  the  degree of charge inversion of the sphere. 
For instance, at the transition point from phase F to A and for $\tilde L =16.3$, the charge of the sphere is effectively reversed, i.e., $\Gamma\simeq 1$ or $\tilde U\simeq 2\tilde Z$. 

Another remarkable point is that $\tilde U$ shows a linear 
 dependence on the sphere charge in phases F and A, indicating that the wrapping process of
 the chain arm(s) occurs locally and continuously in each phase, ensuring an almost constant 
rescaled  net charge, $\tilde U-\tilde Z$, for the complex. This process is  saturated in the fully wrapped phase R. 
Note also that for the longer chain with $\tilde L = 22.4$, 
the transition to phase R takes place at a larger value of the sphere charge not shown in the figure.

\subsection{Role of the chain length} 
\label{sec:upd_effL}

For very small chain length, one can argue that the chain must be completely adsorbed on the sphere
and adopt a fully symmetric configuration (phase F). In this case, the chain forms 
a circular arc which in fact ensures maximal distance between charged segments of the chain and thus minimal chain 
self-repulsion. This situation is visualized in Fig. \ref{fig:upd_Lgal} (top left box) for the rescaled  chain length 
$\tilde L =5.3$, which is somewhat smaller than the circumference of the great circle $2\pi$ (in rescaled units).  
The end-point effects come into play for larger chain length and cause out-of-plane deflection of the end segments 
resulting thus in phase R with broken mirror symmetry (see, e.g., the results for  
$\tilde L = 6.1$ in the figure).  
By further increasing the chain length, a sequence of conformational changes take place and 
the chain eventually unwraps as its  self-repulsion dominates the chain-sphere attraction. For intermediate
values of  $\tilde L$, we find a (discontinuous) transition from phase R to the asymmetric phase A and for very large $\tilde L$, another (discontinuous) 
transition from phase A to a novel expanded conformational phase R$'$ which,  unlike the standard fully-symmetric 
expanded state F, possesses only a two-fold rotational symmetry and exhibits a {\em partially wrapped} state; we thus refer to it as {\em expanded non-mirror-symmetric phase}.
 In the figure, we show two coexisting states at the transition point $\tilde L =22.4$ between the two phase A and R$'$. 
For even larger chain contour length, the Coulomb self-repulsion contribution dominates again and the sequence of phases 
terminates again in the expanded-chain state F (not shown).

\begin{figure}[t]
\begin{center}
\includegraphics[angle=0,width=8.5cm]{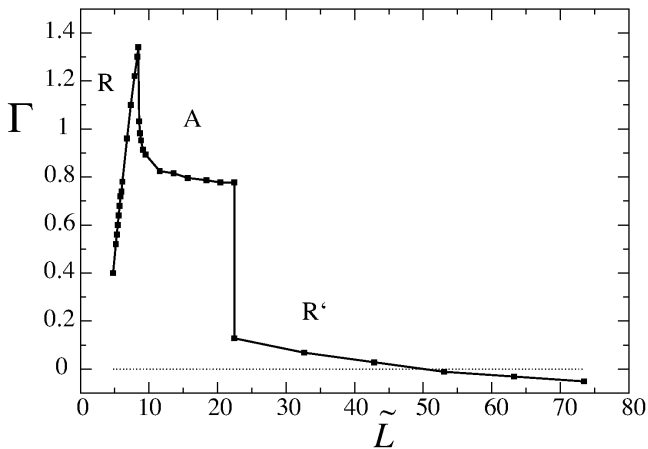}
\caption{Degree of charge inversion of the sphere $\Gamma$ is plotted as a function of $\tilde L$ for the parameter values and configurations shown in Fig.  \ref{fig:upd_overL}.}
\label{fig:upd_overL}
\end{center}
\end{figure}

\begin{figure*}[t]
\begin{center}
\includegraphics[angle=0,width=12cm]{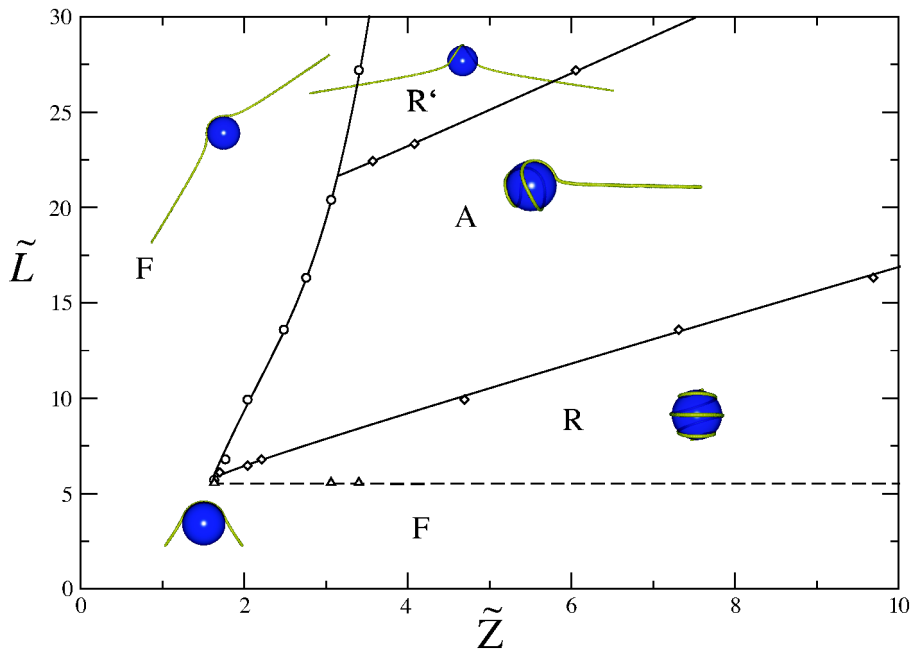}
\caption{Phase diagram displaying different conformational phases of a charged chain wrapped around an oppositely charged nano-sphere and the transitions between these phases in terms of the 
rescaled sphere charge $\tilde Z = Z/(\tau R_{\mathrm{s}})$ and the rescaled chain length $\tilde L=L/R_{\mathrm{s}}$ in the absence of salt
($\tilde \kappa =0$) and for a mechanically  flexible chain ($\tilde l_{\mathrm{p}}=0$). Solid (dashed) lines indicate a discontinuous (continuous) transition at the phase boundaries.}
\label{fig:upd_ZLphase}
\end{center}
\end{figure*}

The  degree of charge inversion, $\Gamma$, shows pronounced variations as these conformational changes occur 
by  increasing the chain length (see  Fig. \ref{fig:upd_overL}). It increases sharply before the transition to phase A takes place (reaching
a maximum value of $\Gamma\simeq 1.4$ in phase R at about $\tilde L \simeq 8.5$) and then drops to a plateau-like region.  
As seen, $\Gamma$ shows pronounced variations 
until the transition to phase R$'$ takes place (at about $\tilde L \simeq 22.4$), where 
$\Gamma$ jumps down  to a much smaller value and slowly decreases to even negative values  due 
to the unwrapping behavior of the chain. 

\subsection{Phase diagram: Chain length vs. sphere charge} 
\label{sec:upd_ZLphase}

In the preceding Sections, we explored various aspects of the conformational changes of a charged chain wrapped 
around a nano-sphere by focusing on the special case of a mechanically flexible chain with $l_{\mathrm{p}}=0$ (i.e., 
with no intrinsic bending rigidity but only an effective one generated by its electrostatic self-repulsion) in the absence of salt screening  
$\kappa =0$. We span various regions of the parameters space by changing the only remaining control parameters, i.e., the (rescaled)
sphere charge $\tilde Z= Z/(\tau R_{\mathrm{s}})$ and the (rescaled) chain length $\tilde L=L/R_{\mathrm{s}}$. The results are   summarized in a 
phase diagram 
in terms of these two parameters as shown in Fig.  \ref{fig:upd_ZLphase}. Different conformational phases (i.e., F, A, R, and R$'$)
are indicated on the figure and the open circles show the transition points between them and the interpolating 
lines represent the corresponding phase boundaries. Such a phase diagram is clearly {\em universal} in that it is valid irrespective of the actual value 
of the chain linear charge density, $\tau$, and the sphere radius, $R_{\mathrm{s}}$.  
Note that in actual units  the graph shown in Fig.  \ref{fig:upd_ZLphase} spans chain length values up to $L = 150$~nm 
and sphere charge values up to  $Z  = 300$ if we adopt the DNA-histone parameters as $R_{\mathrm{s}} = 5$~nm and $\tau = 5.88$~nm$^{-1}$.

As seen,  both for sufficiently small chain length and sufficiently small sphere charge, the ground-state
configuration is given by the fully symmetric phase F which is characterized by the dominant self-repulsion 
of the chain segments.  Short chains in phase F (i.e., for $\tilde L < 5.57$ in the bottom left region of the diagram) are 
highly adsorbed on the sphere, while longer chains take an expanded conformation with two open arms (left margin of the diagram). 
As the chain length is increased at fixed $\tilde Z$, we find the sequences of conformational phases as discussed before, i.e., 
the fully wrapped phase R for small to intermediate values of the chain length, the asymmetric phase A for intermediate to large chain lengths, 
and the expanded non-mirror-symmetric phase R$'$ for even larger values of the chain length. 
It is important to note that these phases appear only 
for sufficiently large sphere charge determined by the phase boundary line between these phase and phase F on the left 
margin of the diagram along the vertical axis. The absolute minimum value of $\tilde Z$ to achieve a configuration
in phase R or A is  $\tilde Z > 1.63 $ (corresponding to   $Z> 48 $ for
the DNA-histone system with $R_{\mathrm{s}} = 5$~nm and $\tau = 5.88$~nm$^{-1}$). 
The largest  degree of charge inversion as discussed previously is obtained  at the transition line between 
phases R and A, which is an experimentally accessible region. 
It is also interesting to note that the boundary line between phases R and A (where the two-fold rotational symmetry is spontaneously
broken) and that between phases A and R$'$ (where the two-fold rotational symmetry is restored again) exhibit a linear dependence 
 for $\tilde L$ as a function of $\tilde Z$ along these lines. 

The phase boundary line between phases F and R (the dashed horizontal line $\tilde L = 5.57$)
represents a {\em continuous} transition, while other phase boundaries (shown by solid lines) represent  {\em discontinuous}
transitions between different phases. The nature of these conformational transitions are reflected in the behavior of the energy 
and other quantities such as the  degree of charge inversion as noted before. However, since these transitions are accompanied
by changes in the symmetry class of the optimal conformation of the chain-sphere complex, it is more convenient to make
use of ``order parameters" that can identify such symmetry changes. Following Refs.  \cite{Kunze1,Kunze2}, we use the order parameters
 \begin{equation}
  \eta  = \frac{{\mathbf r}^2(0) - {\mathbf r}^2(L)}{L^2}, 
\end{equation}
where ${\mathbf r}(0)$ and  ${\mathbf r}(L)$ give the positions of the chain end-points (for  $s=0$ and $s=L$, respectively), and 
\begin{equation}
  \sigma = \frac{1}{L} \int_0^L {\mathrm{d}}s\, \omega(s), 
\end{equation}
where $\omega(s)=\dot {\mathbf n}(s)\cdot{\mathbf b}(s)$ is the so-called torsion associated with the chain contour line, 
${\mathbf r}(s)$, with the vectors ${\mathbf n}(s)$ and ${\mathbf b}(s)$ being defined
using the tangent vector ${\mathbf t}(s)  =\dot {\mathbf r}(s)$ as 
\begin{equation}
	{\mathbf n}(s)  =\frac{\dot {\mathbf t}(s)}{|\dot {\mathbf t}(s)|},\,\,\,\,\,{\mathbf b}(s) = {\mathbf t}(s)\times {\mathbf n}(s), 
\end{equation}
for non-vanishing curvature $|\ddot{\mathbf r}(s)|$ (note also that $|\dot {\mathbf r}(s)|=1$ is assumed within the present model).

\begin{figure}[t]
\begin{center}
\includegraphics[angle=0,width=8.8cm]{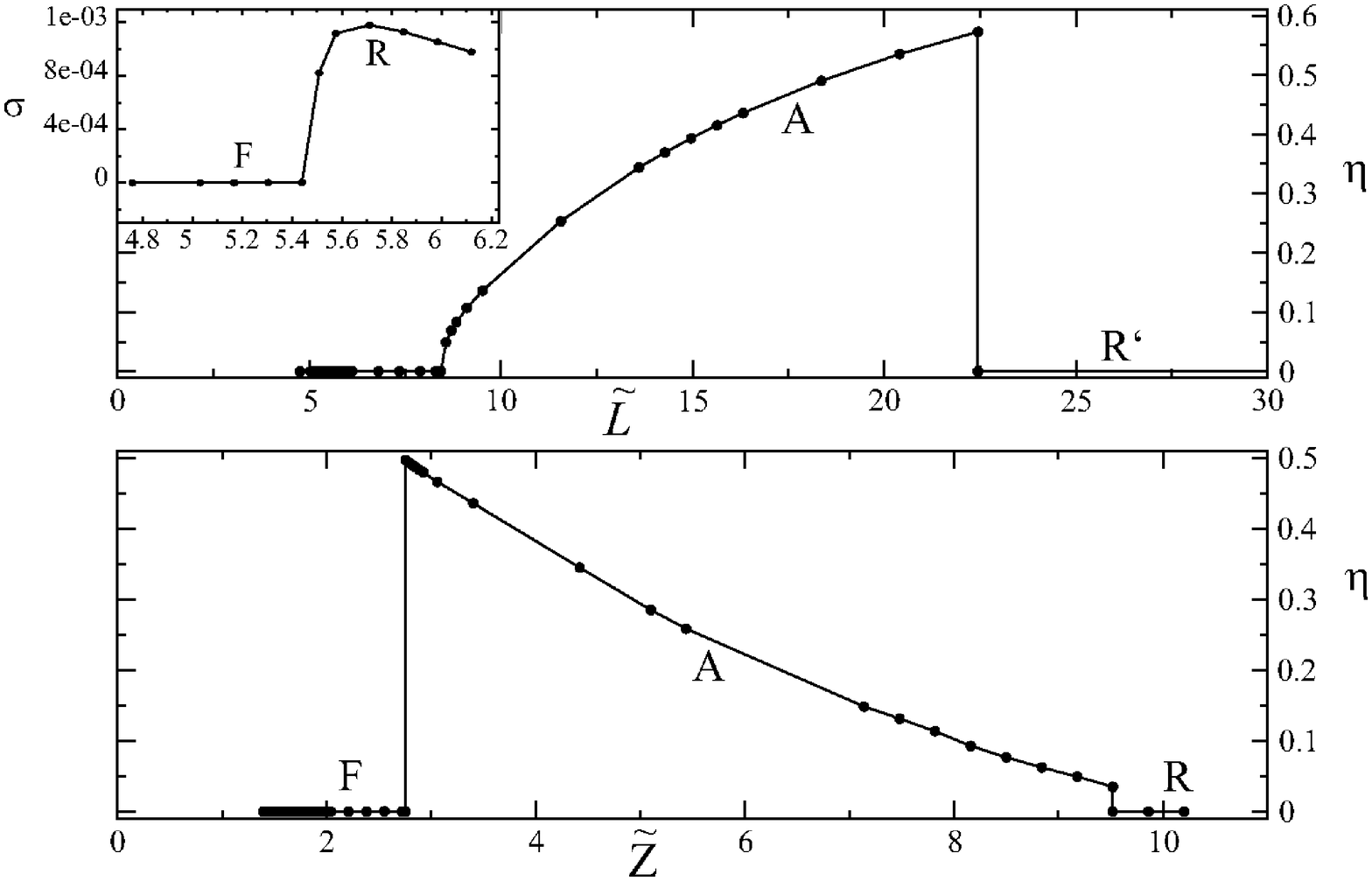}  
\caption{The conformational order parameters for the case with $\tilde \kappa=0$ and $\tilde l_{\mathrm{p}}=0$.  
Top panel shows $\eta$ as a function of $\tilde L$ for $\tilde Z = 3.4$ (corresponding to configurations in Fig. \ref{fig:upd_Lgal}). The inset shows the order parameter $\sigma$ as a function of $\tilde L$ for the same parameters but 
in the vicinity of the continuous transition from phase F to R. 
Bottom panel shows $\eta$ as a function of $\tilde Z$ for $\tilde L = 16.3$ (corresponding to configurations in Fig. \ref{fig:upd_zlgal}).}
\label{fig:upd_ZLetaxi}
\end{center}
\end{figure}

\begin{figure*}[t]
\begin{center}
\includegraphics[angle=0,width=15.cm]{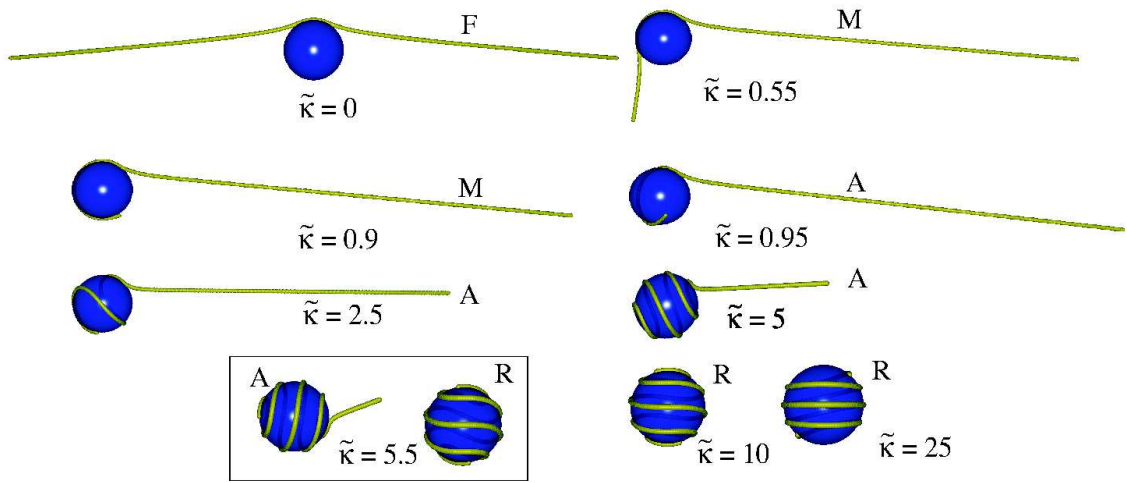}
\caption{Optimal configurations obtained for $\tilde Z = 1$, $\tilde L = 22.4$, $l_{\mathrm{p}} = 0$ and various salt screening parameters 
increasing from $\tilde \kappa=0$ up to $\tilde \kappa=25$ as shown on the graph. The symmetry class of the chain conformation 
(F, M, A or R) is also indicated for each configuration. 
The two coexisting configurations obtained at $\tilde \kappa =5.5$ are shown together in a closed box. For the sake of presentation, some of the  configurations are  enlarged to show the details more clearly.}
\label{fig:upd_kapagal}
\end{center}
\end{figure*}

The order parameter $\eta$ can quantify the rotational symmetry of the chain conformation: it vanishes if an axis through the origin can be found around which 
the chain configuration has a two-fold rotational symmetry and otherwise takes a non-zero value (note that, strictly speaking,   
$\eta =0$ does not necessarily imply a two-fold rotational symmetry for a general  conformation of the chain, but only for the 
ground-state conformation as it has been explicitly checked \cite{Kunze2}). 
The order parameter $\sigma$ can be used as a  measure for 
the deviations of the chain configuration  from a planar (two-dimensional)  conformation: 
If $ \omega(s) =0$ for all $s$, the chain conformation is planar and lies in the plane of mirror symmetry; for a non-planar (three-dimensional) 
ground-state conformation, $ \omega(s) \neq 0$  for some $s$  (again there may be three dimensional chain conformations with $\sigma \neq 0$ 
which possess mirror symmetry but they do not correspond to ground-state configurations \cite{Kunze2}).

In Fig. \ref{fig:upd_ZLetaxi}, we show the changes in the order parameter $\eta$ along the two perpendicular lines $(\tilde Z=3.4, \tilde L)$ (top panel) and 
 $(\tilde Z, \tilde L=16.3)$ (bottom panel)  scanning the phase diagram across regions where the ground-state configurations coincide with those shown 
 in Figs. \ref{fig:upd_Lgal} and \ref{fig:upd_zlgal}, respectively. As seen, the transition from the fully symmetric phase F to asymmetric phase A is indeed associated 
 with a discontinuous jump in $\eta$ (bottom panel). This parameter exhibits a less pronounced jump where another conformational transition takes place to 
 the phase R$'$ (top panel) or phase R (bottom panel); this is because although the chain conformation is three dimensional, its two ends show
 only a small out-of-plane deviation. The continuous transition from phase F to phase R, on the other hand, is not detected by $\eta$ at all (top panel). 
 It can be located by examining the changes in the order parameter $\sigma$ (inset, top panel). 
 
An important point to be noted here is that no mirror symmetry phase 
is  observed in the present phase diagram. This is because in the absence of salt screening, 
the mirror symmetry and the two-fold rotational symmetry of the ground state are found to be broken {\em simultaneously}, 
and they can be decoupled to a sequence of individual symmetry-breakings (thus giving rise 
to an intermediate mirror symmetric phase between  phases F and A)  when there is salt screening in the system 
as we show in the following Section. 

\section{Effects of Salt Screening}
\label{sec:phase_behavior_salt}

We now turn our attention to the case where the system contains a finite amount of added salt (electrolyte) that modifies 
the long-ranged  (bare) form of Coulomb interactions to a short-ranged (screened) DH form; hence, the Hamiltonian 
has the form shown in Eq. (\ref{eq:upd_H_full}) with a finite inverse Debye screening length $\kappa$. In order
to bring out the effects of salt screening more clearly, we shall again simplify the problem by taking a charged chain of 
no intrinsic (mechanical) bending rigidity, $l_{\mathrm{p}} = 0$. The chain stiffness and unwrapping from the sphere 
is thus driven again by no mechanism other than the electrostatic self-repulsion of the chain.

In Fig.  \ref{fig:upd_kapagal}, we show numerically obtained optimal (ground-state) conformations of a charged chain with rescaled 
length $\tilde L = 22.4$  on a nano-sphere of rescaled charge  $\tilde Z= 1$ for various salt screening parameters $\tilde \kappa$.
In actual units and assuming $R_{\mathrm{s}} = 5$~nm and $\tau = 5.88$~nm$^{-1}$, these parameters correspond
to $Z= 29.4$ and $L=112$~nm (equivalent nearly to the length of 330 base pairs of DNA). 
  
The first conformation shown in the figure (top left) corresponds to the case with zero salt discussed  in the
previous Section; it shows an expanded conformation belonging to phase F because of the dominant (unscreened) self-repulsion of 
chain segments.  
As the salt concentration is increased, the chain-sphere attraction become increasingly more relevant  and attracts the two open arms of
the chains more strongly towards the sphere (not shown in the figure) but leaves  the geometrical symmetry of the chain conformation 
 intact until the threshold value $\tilde \kappa = 0.5$ is reached. At this point, the two-fold rotational symmetry is broken spontaneously 
 and the chain adopts a {\em mirror symmetry phase} M. In this phase, one of the chain arms grows in length at the
expense of  the other arm (as if the chain ``slides" or is ``pulled" from one end); the shorter arm is then adsorbed more strongly
on the sphere, while the chain conformation still remains planar or two dimensional. This process clearly diminishes 
the electrostatic repulsion between the two arms. Further increase in the salt concentration leads to an asymmetric conformation A 
as the wrapped end of the chain deviates from its symmetry plane. The degree of chain wrapping obviously increases further in this phase
and the free end of the chain is pulled further upon the sphere as the salt concentration is increased. 
As seen, the number of turns increases
as long as there are enough free segments to be adsorbed. Eventually, a highly {\em ordered helical structure}
 is formed and the complex exhibits a transition to the fully wrapped phase R at $\tilde \kappa = 5.5$. 
For $R_{\mathrm{s}}=5$~nm, this value corresponds to $\kappa = 1.1$~nm$^{-1}$, which is comparable to 
the physiologically relevant regime of about 100~mM monovalent salt. 
For a mechanically flexible chain with $l_{\mathrm{p}}=0$,  phase R remains stable as the salt screening is increased further. 
The presence of a finite persistence length can cause chain unwrapping at high salt concentrations where all electrostatic interactions 
(including the chain-sphere attraction and the electrostatic screening of the chain) 
are completely screened. We shall return to the effects of the persistence length later 
in Section \ref{sec:phase_behavior_ellp}. 
 
\begin{figure}[t]
\begin{center}
\includegraphics[angle=0,width=8.5cm]{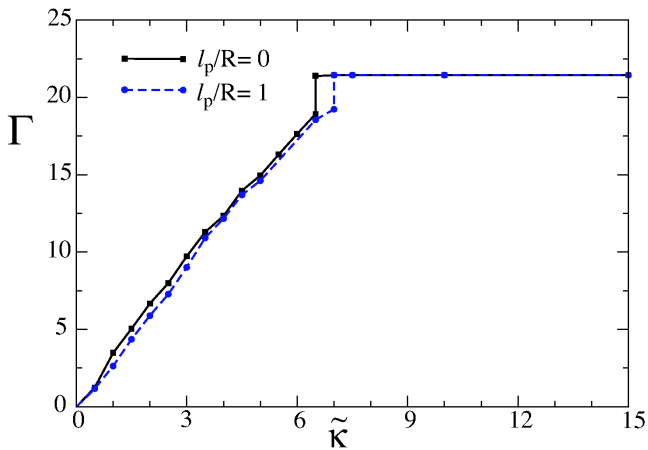}
\caption{Degree of charge inversion of the sphere $\Gamma$, Eq. (\ref{eq:upd_gamma}), is shown as a function
of $\tilde \kappa$ for the rescaled sphere charge $\tilde Z =1$,  rescaled chain length $\tilde L = 22.4$ and two different values of the mechanical 
persistence length $l_{\mathrm{p}} = 0$ and $l_{\mathrm{p}}/R_{\mathrm{s}}=1$.}
\label{fig:upd_overchargingkapa}
\end{center}
\end{figure}

\begin{figure*}[t]
\begin{center}
\includegraphics[angle=0,width=12cm]{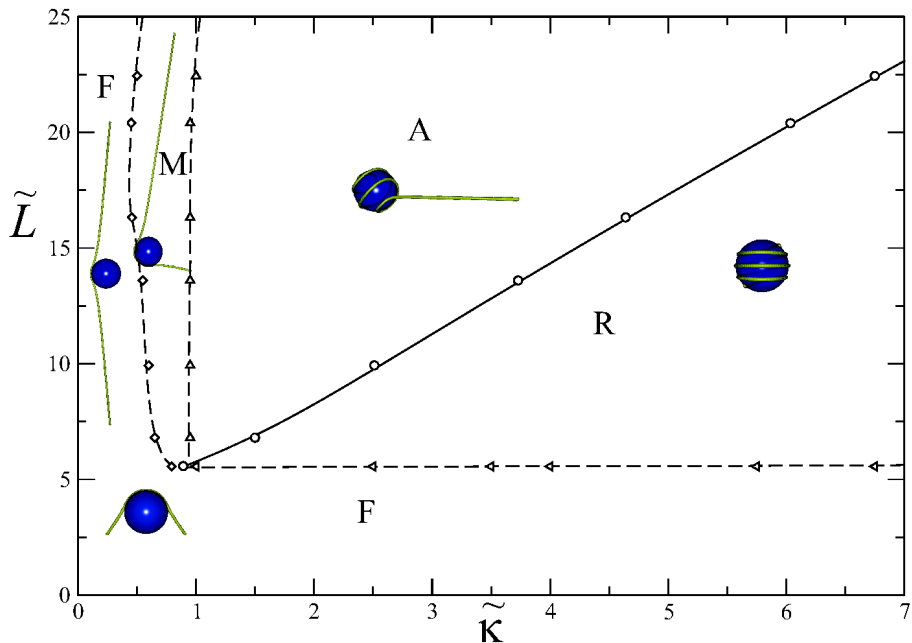}
\caption{Phase diagram displaying different conformational phases of a charged chain wrapped around an oppositely charged nano-sphere and the transitions between these phases in terms of the 
rescaled salt screening parameter $\tilde \kappa = \kappa R_{\mathrm{s}}$ and the rescaled chain length $\tilde L=L/R_{\mathrm{s}}$  for a mechanically  flexible chain ($\tilde l_{\mathrm{p}}=0$) and fixed rescaled sphere
charge $\tilde Z = 1$. Solid (dashed) lines indicate a discontinuous (continuous) transition at the phase boundaries.}
\label{fig:upd_kapaL}
\end{center}
\end{figure*}

An interesting feature in the fully wrapped phase R is that the helical structure becomes increasingly more compact 
(with smaller lateral spacing between two successive turns of the helix) 
as $\tilde \kappa$ is increases. Consequently, 
this compact chain wrapping can result in a highly charge-inverted complex with an effective charge of 
up to an order of magnitude larger than the bare charge of the sphere as shown 
in Fig.  \ref{fig:upd_overchargingkapa} (see symbols connected by a solid line for $l_{\mathrm{p}}=0$). 
The linear increase of  the charge inversion degree $\Gamma$ with $\kappa$ again reflects the progressive 
wrapping of the chain around the sphere until it  saturates to a constant value in phase R, which is equal to $\Gamma\simeq 21.5$ in
the figure. Such a large charge inversion is possible only in the presence of strong salt 
screening effects, where electrostatic interactions are very short-ranged  \cite{Netz2,Netz99a,Netz1,Schiessel03_rev} (compare this figure with
Figs. \ref{fig:upd_overZ} and \ref{fig:upd_overL}).
Note that the variation of the chain length in this case  (data not shown) only affects the saturation threshold  
of $\Gamma$; the latter shifts to a smaller (larger) salt screening parameter for a smaller (larger) chain length, while
the slop of the linear increase and other features of the graph remain unchanged. 

\subsection{Phase diagram: Chain length vs. salt screening parameter} 
\label{sec:upd_kapaLphase}

The interplay between salt screening effects and the chain length can be summarized in a phase diagram as shown 
in Fig. \ref{fig:upd_kapaL}. Here we have fixed $\tilde Z=1$ but the analysis can be repeated for different values of $\tilde Z$
(or, alternatively, one can obtain a $\tilde \kappa-\tilde Z$ phase diagram for a given length of the chain to supplement the present 
phase diagrams, which has indeed been done before  \cite{Netz1,Kunze1,Kunze2} and we shall not discuss it further here). 
Note that in actual units and assuming the DNA-histone
parameters as $R_{\mathrm{s}} = 5$~nm and $\tau = 5.88$~nm$^{-1}$, the 
phase diagram in Fig. \ref{fig:upd_kapaL} covers chain length values up to $L=125$~nm and  salt screening parameters 
up to $\kappa = 1.4$~nm (equivalent to around 200 mM of monovalent salt concentration). 

The stable (ground-state) configurations that are obtained from our numerical results are again classified in terms of their 
conformational symmetry phases and the nature of the transition lines between these phases is determined 
from the appropriate order parameters (or the conformational energy of the complex).  

As discussed in detail before, the short-chain regime $\tilde L < 5.57$ is dominated by the fully symmetric 
phase F for the whole range of $\tilde \kappa$, while the regime of intermediate and large chain lengths  shows more complex structures 
corresponding to symmetry phases R, A and M. The fully wrapped and the asymmetric phases R and A are 
stable for larger values of $\tilde L$ but only for  intermediate to large salt concentrations, specifically, 
above the rescaled screening parameter  $\tilde \kappa = 1$ (equivalent to the actual value of $\kappa=0.2$~nm$^{-1}$ for $R_{\mathrm{s}}=5$~nm). 
The conformational transition from phase F to phase R turns out to be a {\em continuous} one (dashed line). 
The fully wrapped phase R is found as the stable phase at the intermediate regime of chain lengths, while the asymmetric phase A 
is  stable for large chain lengths. These two phase are separated by a {\em discontinuous} transition line (solid line). For very large chain lengths (not shown), the self-repulsion of the  chain becomes dominant causing an unwrapping transition from phase A back to   
the phase F again.

\begin{figure*}[t!]
\begin{center}
\includegraphics[angle=0,width=12.cm]{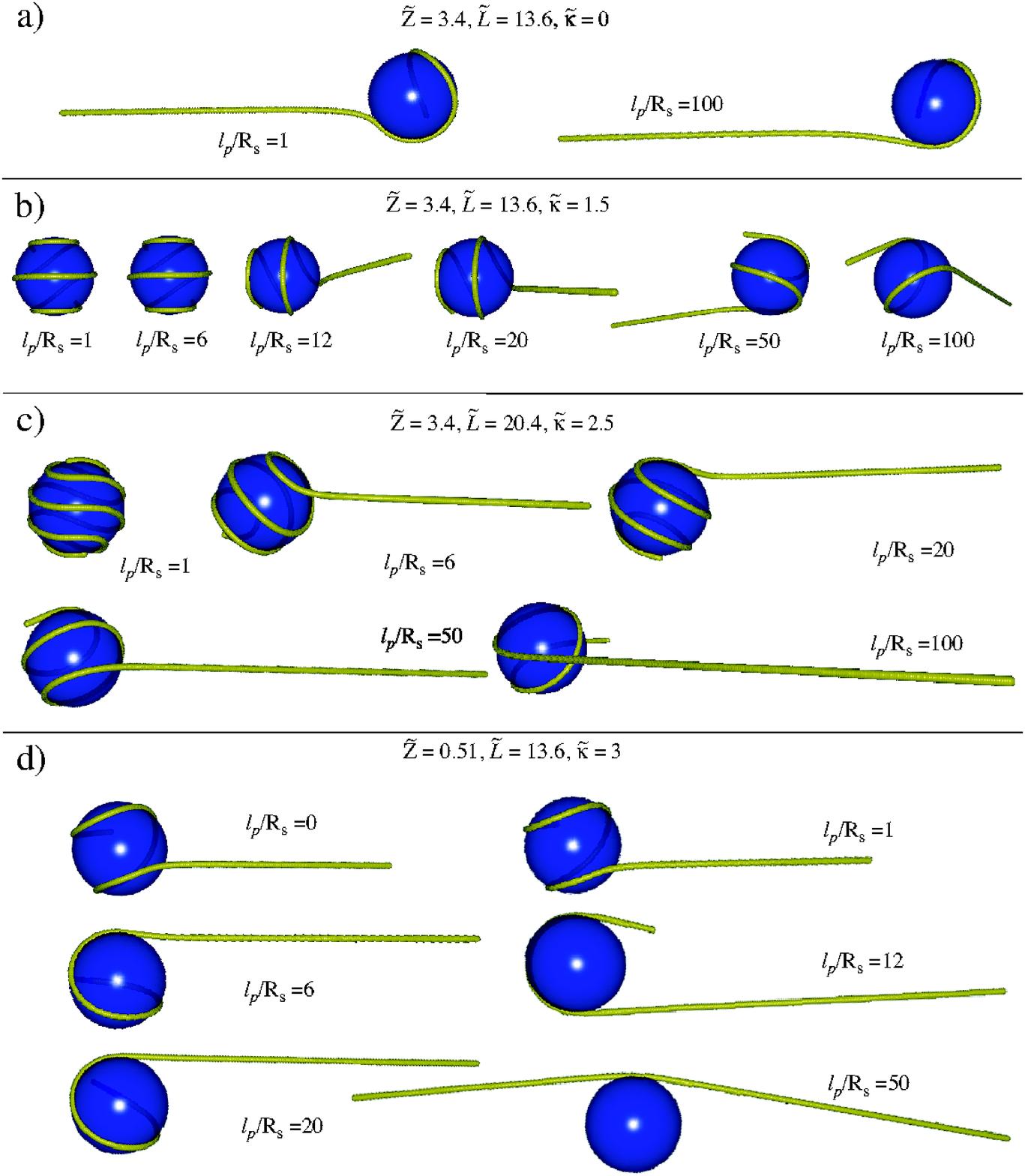}
\caption{Optimal configurations obtained by changing the  mechanical persistence length, $l_{\mathrm{p}}$, of the charged chain wrapped around
an oppositely charged nano-sphere. The parameter values  for each case are shown on the graph. }
\label{fig:upd_lp}
\end{center}
\end{figure*}

For very small salt screening parameters, the electrostatic interactions are long-ranged and thus the 
conformational phase F dominates for the whole range of $\tilde L$ (left margin of the graph). As $\tilde \kappa$ is increased,  
the two-fold rotational symmetry and the mirror symmetry are broken one at a time before the asymmetric phase
A is reached: first, the two-fold rotational symmetry is broken leading to a continuous transition to a narrow
region characterized by the mirror symmetry class M and then this latter symmetry is broken engendering
another continuous transition to the phase A. 
Therefore, for a sufficiently large chain length $\tilde L > 5.57$ (corresponding to
$L= 27.85$~nm for $R_{\mathrm{s}}=5$~nm, or equivalently, about the length of 82 base pairs of DNA), one can capture 
all four symmetry classes. 

An interesting feature is that the phase boundaries connected with phase M 
show little dependence on the salt concentration, in contrast to those between phases A and R.


\section{Effects of Mechanical Stiffness }
\label{sec:phase_behavior_ellp}

So far we have focused on the case of  a charged polymer chain with no mechanical persistence length ($l_{\mathrm{p}}=0$). 
The mechanical bending stiffness 
 favors unwrapping of the chain from the oppositely charged nano-sphere, and thus competes with 
the electrostatic self-repulsion of the chain. Our foregoing
analysis therefore gives the role of electrostatic interactions in the wrapping-unwrapping behavior of the chain 
conformation and in absence of any elastic contributions. It should be noted however that the effects of mechanical and electrostatic chain stiffness can be qualitatively
similar (and can be even put together using the notion of effective or renormalized persistence length  \cite{netz-orland,Skolnick77,Odijk,Fixman,Odijk95,Khokhlov})  but also that,
for highly charged chains or small salt screening parameters, the mechanical persistence length is subdominant and
leads to no qualitative changes in our results from those presented in the previous Sections. This can be seen from the optimal configurations for zero salt concentration $\kappa=0$
in Fig. \ref{fig:upd_lp}a, where  the chain conformation is found to be affected very little when $l_{\mathrm{p}}$ is increased by two orders of
magnitude from $l_{\mathrm{p}}/R_{\mathrm{s}}=1$ up to $l_{\mathrm{p}}/R_{\mathrm{s}}=100$.

The mechanical persistence length of the chain starts to play a role when the salt concentration is finite and sufficiently large. In Figs. 
 \ref{fig:upd_lp}b-d, we show a selection of significant configurational changes in the presence of salt screening 
as the mechanical persistence length changes. 

In Fig. \ref{fig:upd_lp}b, the sphere charge and chain length are the same as in Fig.  \ref{fig:upd_lp}a and are fixed at intermediate values  $\tilde Z = 3.4$ and 
$\tilde L =13.6$, and  the inverse screening length is  $\tilde \kappa = 1.5$  (corresponding to $Z=100$, $L=68$~nm  and $\kappa =  0.3$~nm$^{-1}$
for the DNA-histone system). As seen, the fully wrapped conformation of the chain (phase R) remains intact up to $l_{\mathrm{p}}/R_{\mathrm{s}}=6$, which 
is a rather large persistence length as compared with the sphere radius (in actual units and with  $R_{\mathrm{s}}=5$~nm it gives  $l_{\mathrm{p}}=30$~nm, 
which coincides with the bare mechanical persistence length of DNA \cite{Sobel,Frontali,Rief99,Hagerman}). As $l_{\mathrm{p}}$ is further increased beyond this point, a discontinuous transition to the asymmetric phase 
A takes places; this latter phase persists (although the chain gradually unwraps) even up to $l_{\mathrm{p}}/R_{\mathrm{s}}=100$ (very stiff chain), reflecting the fact that the chain-sphere
 attraction is still the dominant factor. 
A similar trend is seen  in Fig. \ref{fig:upd_lp}c where we have the same sphere charge but a  larger salt screening  parameter ($\tilde \kappa =2.5$) and 
a larger chain contour length  ($\tilde L =20.4$); this figure clearly shows  the process in which the mechanical chain
stiffening gradually unwraps the  chain from an initially highly wrapped (helical) conformation  by reducing the number of chain turns
around the sphere.  

To demonstrate the complete unwrapping transition, we take a smaller sphere charge of $\tilde Z = 0.51$
and a larger salt concentration of $\tilde \kappa =3$  in Fig. \ref{fig:upd_lp}d (in actual units,we have  $Z=15$ and $\kappa = 0.6$~nm$^{-1}$
for the DNA-histone system). 
In this case, the chain adopts an asymmetric conformation (phase A) at small $l_{\mathrm{p}}/R_{\mathrm{s}}$ and after a sequence of conformational changes 
exhibits a transition to the mirror symmetric phase M  (see the configurations for  $l_{\mathrm{p}}/R_{\mathrm{s}}=12$ and 20) and eventually 
the expanded phase F (see the configuration for $\tilde l_{\mathrm{p}}=50$).

\section{Conclusion and Discussion}
\label{sec:discussion}

We investigated the structural phase behavior of complexes formed by a charged polymer chain and 
an oppositely  charged nano-sphere using a  primitive chain-sphere model and extensive numerical optimization
methods. In this case, the chain conformation is determined 
by a competition between the chain elasticity and the electrostatic  repulsion between the
chain segments, which both tend to unwrap the chain from the sphere, and the electrostatic attraction between the chain 
and the sphere, which tends to wrap the chain around the sphere. These interactions are accounted for via an effective Hamiltonian, which
is then minimized  with respect to  the whole conformation of the polymer chain in order to obtain the optimal chain conformation
for a given set of parameters (such as the contour length, the persistence length and the charge density of the polymer, the radius and charge of the sphere and the  
salt concentration in the bathing solution). Our study thus applies to strongly coupled complexes where thermal fluctuations are negligible and
the energetic ground state is dominant.

We employed a rescaled representation in order to obtain generic results 
that can be applied to a large variety of charged chain-sphere complexes. In other words, our results map a wide range of actual system
parameters  to a small 
set of dimensionless control parameters including  rescaled chain persistence length,  rescaled chain length,  rescaled 
sphere charge and  rescaled salt concentration. The previously studied case of DNA-histone complexes \cite{Kunze1,Kunze2,borujerdi,borujerdi2,borujerdi3,Netz99a} 
within the context of nucleosome core particles is thus included as a special case. We have  explored various
aspects of the conformational changes that may be engendered by changing  the control parameters and thus determined
the global phase behavior of such complexes. 

We showed that variations in these parameters can trigger 
a wrapping transition in a number of different ways. 
For instance, by increasing the sphere charge for an intermediate range of chain lengths we find a transition from an expanded state 
to a fully wrapped state in which the sphere charge is overcompensated  by excessive adsorption of the chain segments.
But this can occur only if the chain has a minimum  rescaled  length of $\tilde L=L/R_{\mathrm{s}}>5.57$  and the sphere 
has a minimum charge of 
$\tilde Z = Z/(\tau R_{\mathrm{s}})=1.63$ in the situations where one deals with  a mechanically flexible chain
in the zero-salt limit. 
By increasing the salt concentration,  the wrapped state becomes stable 
in a wider range of chain lengths. Note that increasing the chain length in general tends to  unwrap the chain because of its stronger electrostatic 
self-repulsion. At elevated salt concentration, electrostatic interactions are screened 
and one reaches at a highly compact wrapped state, where as shown, the charged chain may wrap around
the sphere in several complete turns. This gives rise to a large charge inversion of the complex with a net
 charge that may exceed the bare  charge of the sphere by more than an order of magnitude. 
The wrapped phase can occur again with a minimum chain length and salt screening parameter (e.g., for fixed $\tilde Z=1$ and for 
a mechanically flexible chain, these minimum values are given by $\tilde L=L/R_{\mathrm{s}}=5.57$ and $\tilde \kappa  = \kappa R_{\mathrm{s}}>1$, respectively). 
The wrapped phase is thus predicted to occur at intermediate salt concentrations or sufficiently large sphere charge in agreement with previous studies 
\cite{Kunze1,Kunze2,Stoll01b,Stoll02,Stoll11,Schiessel,Wallin96_I,Linse,Jonsson01a,Jonsson01b} and also recent experiments \cite{Strauss,Haronska,Mohwald,Bielinska,Rieger,Yager1,Yager2,Dieterich, Ausio,Russev,Brown,Libertini1,Libertini2,Libertini3,Oohara,Weischet}.

 Our study therefore generalizes the previous works that were  focused on the specific case of DNA-histone complexes \cite{Kunze1,Kunze2,borujerdi,borujerdi2,borujerdi3,Netz99a} and 
 where a phase diagram for the conformational wrapping of the  chain (with DNA-histone parameters) was obtained in terms of the sphere charge and the salt screening parameter. 
In this work, we  pursued a more general goal by setting out to study the
structural properties of {\em generic} chain-sphere complexes by including 
the most generic elastic and electrostatic aspects of such complexes and disregarding 
specific features that may be realized in any particular application. 
In principle, such details can be included in the chain-sphere model provided that 
they  can be included as additional terms in the Hamiltonian of the system 
or can be represented as physical constraints on the conformation of the chain. The present methods can also be used to study more detailed or sophisticated models as such provided that the thermal fluctuations are not dominant and the chain-sphere complex remains strongly coupled.  

The effects of  thermal fluctuations around the ground-state configurations can be accounted for by means of a systematic saddle-point 
approximation as discussed in Ref. \cite{borujerdi2}. If these effects are strong, one may  enter the regime of weakly charged chain-sphere
complexes which are not considered here (see, e.g.,   \cite{Podgornik3,Sens99,Golestanian,Sens,Stoll01b,Stoll02,Stoll11,Muthu94} and references therein). 


Our model takes advantage of a few other approximations and simplifications that will be discussed in what follows. First, we have neglected the excluded-volume interaction between chain 
segments (note that the chain-sphere excluded-volume interaction is accounted for explicitly via a semi-rigid repulsion and that the 
the finite radius of the chain is accounted for by including it in the effective radius of the sphere).  The intra-chain excluded-volume interactions can play a role 
in the regime where the chain wraps around the spheres more than two turns (e.g., at elevated salt concentrations) and depend on the 
lateral spacing between subsequent turns of the chain. We have also neglected the twist degrees of freedom explicitly which can be present in certain
polymers such as DNA. This is however a good approximation in the present case since the  two 
ends of the chain within our model are free to rotate.  
We have also neglected possible surface features of the charged sphere which may be present in the case of proteins, nano-colloids or
other spherical objects that complex with charged polymers.  This obviously amounts to a additional simplification when our model is applied to such systems since these objects may possess a nontrivial (possibly heterogenous or discrete)  charge pattern  
or surface regions with specific binding sites as is in fact the case for the histone octamers \cite{Schiessel03_rev,The_cell}. In fact, the shape
of the core histone in the case of nucleosome core particles also deviates from a sphere and resembles  a cylindrical, wedge-shaped structure with 
a diameter of about 7~nm and mean height of 5.5~nm. Our model makes use of spherical nano-spheres instead and thus when applied to histone proteins  \cite{Kunze1,Kunze2,borujerdi,borujerdi2,borujerdi3,Netz99a}, we take an effective histone radius of $R_{\mathrm{s}}=5$~nm nearly equal 
to the combined mean radius of the core histone and that of the DNA wrapped around it. The presence of histone tails \cite{Schiessel03_rev} and the dielectric polarization of the spherical core  \cite{Winkler12} are among other effects that we have neglected in our study; the latter can play a role especially when the sphere is impermeable and exhibits a low dielectric constant. 

We should also note that the DH potential employed here  to describe the electrostatic pair interactions
neglects nonlinear electrostatic effects as well as possible correlation effects introduced by 
the ionic atmosphere around charged objects. The latter effects become important only in the presence of multivalent
ions which go beyond the scope of the present work \cite{phys_rep}. The nonlinear effects, on the other hand, may be present in principle for
sufficiently highly charge chains and spheres \cite{Manning69}  although they  are  strongly suppressed for moderately large salt concentrations  \cite{netz-orland}. 
These effects can be incorporated on the mean-field level using 
the so-called nonlinear Poisson-Boltzmann equation \cite{Israelachvili}. This equation is however very difficult to solve in the present geometry 
given the nontrivial shape of the chain conformation but the corresponding nonlinear effects can be included on the (linear) DH level
in an effective way, that is, by replacing the chain linear charge density with a renormalized value. This is a well-explored procedure \cite{Manning69}  
but it should be noted that the rescaled representation used in our study is applicable to any given chain linear charge density and thus, in applying 
our results to actual systems, one can directly adopt the renormalized linear charge density of the chain instead of its bare value.  


\end{document}